\documentclass[runningheads]{llncs}
\usepackage{microtype}
\usepackage{acronym}
\usepackage{url}
\usepackage{textcomp}
\usepackage{xcolor}
\usepackage{subfig}
\usepackage{setspace}
\usepackage[british]{babel}
\usepackage{float}
\usepackage{floatflt}
\usepackage{fancyhdr}
\usepackage{color}
\usepackage{booktabs}
\usepackage[pdftex,bookmarks=true,plainpages=false,pdfpagelabels=true]{hyperref}
\usepackage{mdwlist}
\usepackage{enumerate}
\usepackage{graphicx}
\usepackage{paralist}
\usepackage{array}
\usepackage{longtable}
\usepackage{listings}
\usepackage[utf8]{inputenc}
\usepackage[capitalize, noabbrev]{cleveref}
\usepackage{tabularx}
\usepackage{multirow}
\usepackage{bigstrut}
\usepackage{verbatim}
\usepackage{geometry}

\newcolumntype{L}[1]{>{\raggedright\let\newline\\\arraybackslash\hspace{0pt}}p{#1}}
\newcolumntype{C}[1]{>{\centering\let\newline\\\arraybackslash\hspace{0pt}}p{#1}}
\newcolumntype{R}[1]{>{\raggedleft\let\newline\\\arraybackslash\hspace{0pt}}p{#1}}
\newcommand{\quotes}[1]{``#1''}

\begin{document}
\title{Using Process Models to understand Security Standards - Authors' Copy}
%
%
\author{Fabiola Moy\'on\inst{1}\orcidID{0000-0003-0535-1371}
\and  
Daniel M\'endez \inst{2}\orcidID{0000-0003-0619-6027}  \and   
Kristian Beckers\inst{3} 
\and   
Sebastian Klepper\inst{4} } 
%
\authorrunning{Authors' Copy}
%
\institute{Siemens Technology, Munich, Germany \email{fabiola.moyon@siemens.com}
\and
Blekinge Institute of Technology, Sweden, and fortiss GmbH, Germany \email{daniel.mendez@bth.se}
\and
Social Engineering Academy, Munich, Germany \email{kristian.beckers@social-engineering.academy}
}
\maketitle              
\begin{abstract}
Many industrial software development processes today have to comply with security standards such as the IEC~62443-4-1. These standards, written in natural language, are ambiguous and complex to understand. This is especially true for non-security experts. Security practitioners thus invest much effort into comprehending standards and, later, into introducing them to development teams. However, our experience in the industry shows that development practitioners might very well also read such standards, but nevertheless end up inviting experts for interpretation (or confirmation). Such a scenario is not in tune with current trends and needs of increasing velocity in continuous software engineering. 

In this paper, we propose a tool-supported approach to make security standards more precise and easier to understand for both non-security as well as security experts by applying process models. This approach emerges from a large industrial company and encompasses so far the IEC~62443-4-1 standard. We further present a case study with 16 industry practitioners showing how the approach improves communication between development and security compliance practitioners. 

\keywords{Secure Software Engineering\and Security standards\and Business Process Modeling  \and Industrial Control Systems}
\end{abstract}
\section{Introduction}
Industrial Automation and Control Systems (IACS) are at risk of outages caused by vulnerabilities in information technology. Security risk could be reduced by following a strict secure development life cycle (SDLC)~\cite{Keramati:2008}. For IACS such a secure development process should be compliant with norms like the IEC~62443-4-1 security standard (short: 4-1)~\cite{IEC41:2018}. The 4-1 standard provides a blueprint for a SDLC with the aim of reducing impact on critical infrastructure running IACS systems. This standard is based on SDLC standards such as ISO~27034, BSIMM or Secure CMMI~\cite{ISO27034,BSIMM,SCMMI}.

Compliance with security standards, such as the 4-1 standard, is especially challenging for large-scale industrial companies~\cite{Fitzgerald:2013,Bell:2017}. While they move from traditionally plan-driven development to an agile approach, complexity increases since security standards have to be integrated by several collaborating agile teams. Development practitioners are urged on a daily basis to make decisions on system security~\cite{Turpe:2017,ahola_handbook_2014}, regardless whether these decisions are made implicitly or not. Hence, practitioners need awareness of requirements as outlined by security standards in order to implement them as early as possible in the SDLC (e.g., in every sprint). Moreover, in case of system attacks, development teams need to cooperate immediately and extensively with security experts; therefore, they need prior understanding of the aforementioned standards.
\newline
However, security standards are complex and ambiguous~\cite{Beckers:2015}. Development practitioners find them difficult to understand and usually require interpretation and explanations by security experts. Moreover, to accurately implement standards, security experts are essential to tailor requirements to the existent development process (see~\cite{Maidl:2018}). 

Models are described as useful for overcoming the complexity of security concepts~\cite{Sunkle:2015,Leitner:2013,Riesner:2010,AlHamdani:2009,Hu:2009}, however current contributions do not focus on modeling security standards.

In order to tackle the issue of complex and ambiguous security standards, a process-model-based representation of security standards might contribute to the understanding. 

In this paper, we present an approach to improve understanding of security standards with model-based representations. Our contribution is threefold and aims at bridging current gaps between security compliance and software engineering~\cite{Jaatun:2017}.

First, we contribute \emph{a visual description of the IEC 62443-4-1 security standard}. We present this in the form of process models, following the Business Process Model and Notation (BPMN)~\cite{BPMNguide:2008}. Our target is to present the standard in a more intuitive way and allow easier review, thus, helping development practitioners and security experts to have more focused discussions compared to just textual descriptions.

Second, we conduct an \emph{qualitative evaluation of the IEC~62443-4-1 process models} at Siemens AG, a large industrial company. Given that the security standard is in its introductory phase (released in 2018), the evaluation is still in a preliminary form, focusing particularly on expert interviews. At the same time, we overcome current lacks in reported evidence from contemporary evaluations either having a particularly small sample or invoking students as subjects (c.f.~\cite{Othmane:2017}).

Third, we contribute \emph{tool support for the 4-1 standard process models} with ARIS and Symbio~\cite{ARIS,Symbio}. Both tools support the visualization of the models and improve dissemination within the company. Additionally, tool support leverages change management and accuracy to BPMN notation. 

Our long-term objective is to enable security compliance (including training) for continuous software engineering in industrial environments. By providing a comprehensive representation of the 4-1 standard, its integration into existing agile frameworks is possible.  Besides the qualitative evaluation with practitioners (presented in this paper), we reported already on how the 4-1 process models serve as building blocks to: (a) visualize and close the security compliance gap of the Scale Agile Framework (SAFe) (c.f.~\cite{Moyon:2018}), a process framework used for large scale industrial software development~\cite{safeWeb}, and (b) leverage security compliance assessment with framework and tool (c.f.~\cite{Moyon:2020:tool,Daennart:2019}).

We concentrate this paper on the process models and qualitative evaluation, and we kindly refer the reader to the online material at \url{ https://doi.org/10.6084/m9.figshare.8063159} for further information on the detailed models, meta model and tools. The remainder of this paper is structured as follows: \Cref{sec:modeling} presents an intro to the standard and the modeling work. \Cref{sec:study_design,sec:results} show the study design and results. Finally, we conclude by interpreting results in \cref{sec:conclusion}.

\section{Model Implementation of the 4-1 Standard}
\label{sec:modeling}
The 4-1 standard provides a blueprint for a SDLC considering eight practices namely: \textbf{security management} (SM), treating security activities to implement the product's life cycle; \textbf{specification of security requirements} (SR), referring to accurate eliciting IACS security capabilities; \textbf{secure design} (SD), ensuring security as basis in overall architecture; \textbf{secure implementation} (SI), guiding the use of secure coding techniques; \textbf{security verification and validation testing} (SVV), describing security plans and test types; \textbf{management of security-related issues} (DM), providing a flow to handle security issues; \textbf{security update management} (SUM), providing a delivery flow for security updates; and finally \textbf{security guidelines} (SG), describing the required guidelines to secure deploy the product.

To approach our contribution, the first step is to generate a model representation of the 4-1 standard. Models can support development and security practitioners to better understand and comprehend the norm. As basis for the modeling, we start understanding basic constructs and relationships of the 4-1, as well as the elements of a process model. This helps to determine that process models are suitable to describe the standard. To facilitate comprehension of the models, we use the Business Process Model and Notation 2.0 (BPMN)~\cite{BPMNguide:2008} since BPMN elements are easy to understand for technical and non-technical people. Hence process models will include elements like: tasks (t), events (e), artifacts (a) (c.f. Meta-models file in online material). BPMN elements are enumerated and referred with the abbreviation of the 4-1 practice. Example: SR-a1 states for Security Requirements Practice artifact 1. 

\begin{figure*}[htbp]
	\centering
	\includegraphics[width=0.79\textwidth]{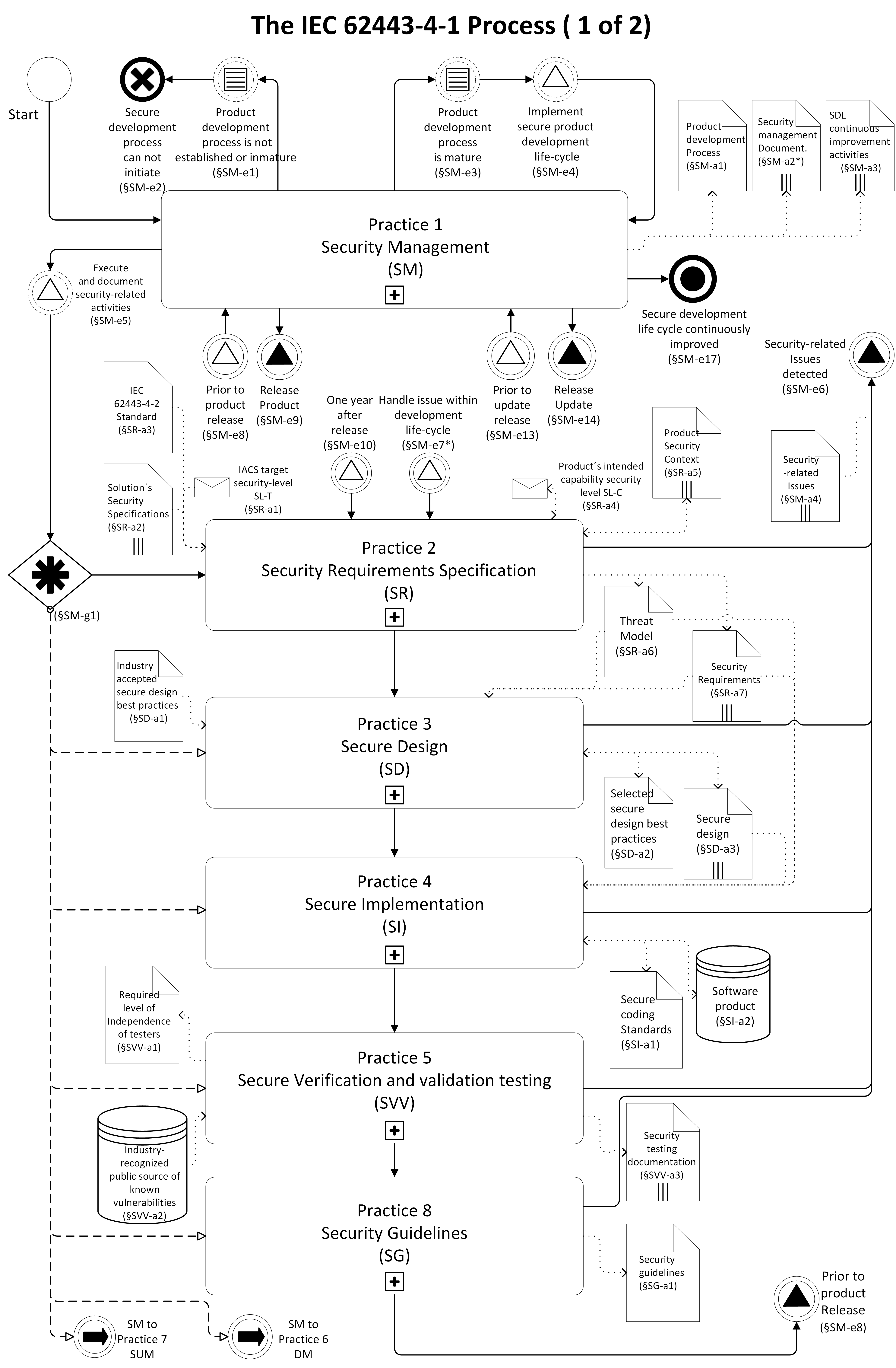}
	\caption{The IEC~62443-4-1 Security Standard Overview Process Model (Excerpt). BPMN diagram depicting the standard practices, their flow, and artifacts.}
	\label{fig:41_global01}
\end{figure*}

Subsequently, we modelled the high-level overview of the 4-1 standard to depict practices and its input/output artifacts, as well as events that trigger further practices. \Cref{fig:41_global01} shows the first half of this view with each 4-1 practice. 

\begin{figure*}[htbp]
	\centering
	\includegraphics[width=0.9\textwidth]{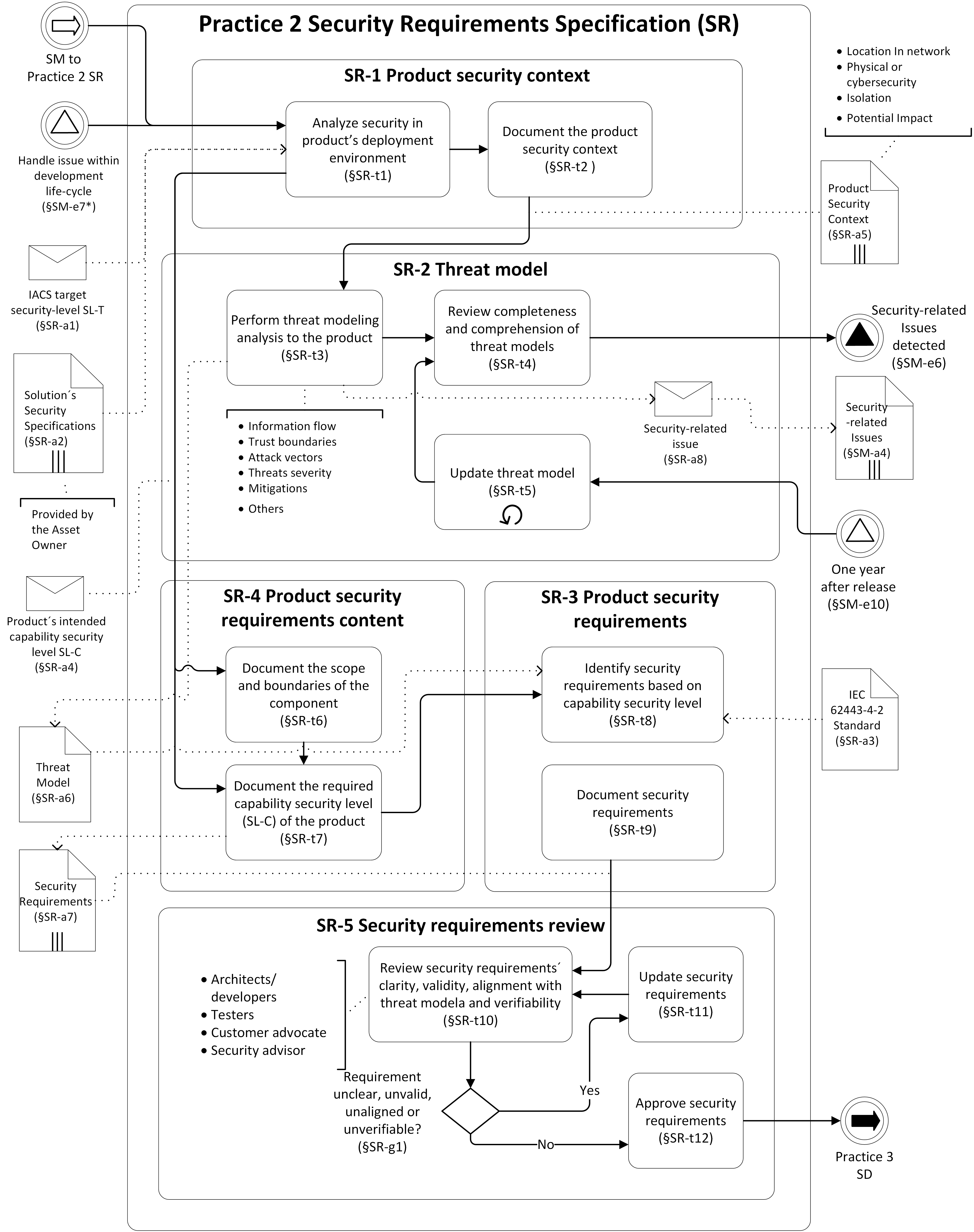}
	\caption{The IEC~62443-4-1 Practice 2 Security Requirements (SR) process model. BPMN diagram depicting the 4-1 standard requirements to document security capabilities in IACS.}
	\label{fig:41_practice2}
\end{figure*}

Besides the overview, each practice is refined into individual models depicting inner tasks and how the input/output artifacts are generated. The model of the 4-1 practice \emph{Security Requirements} (SR) is shown in \cref{fig:41_practice2}. The complete set of process models is available online at the file \emph{The IEC 62443-4-1 Standard Process Models}. During modeling, we continuously validated the outcomes for completeness, correctness, and consistency. Prior the evaluation study, experts on the standard checked for missing, wrong, or misplaced elements as well as connections between tasks, flows, and artifacts.

Finally, in order to improve dissemination and knowledge transfer, we provide the process models in several tool formats. As an initial approach Microsoft Visio was used, allowing easy distribution of the models as PDF files, however, with limitations in change management. This leads us to use BPMN modeling tools. Models are available in: ARIS, an intuitive visualization tool, which provides the necessary flexibility to apply changes with an acceptable learning curve~\cite{ARIS}; and Symbio, that represents artefacts, tasks and events as rectangles with different icons and colors. Even though less flexible than ARIS, it provides more detailed views of models as matrices~\cite{Symbio}. Excerpt of process models in the tools are available online at file \emph{Tool Support}. 


\section{Evaluation Design}
\label{sec:study_design}
We evaluated our 4-1 process models via expert interviews involving 16 practitioners working at Siemens in security compliance or agile software engineering. Among these experts is an IEC committee member for 4-1.

Our overall goal is to investigate the subjective usefulness of our process models to transfer knowledge of security standards, as represented by the IEC~62443-4-1. Through the evaluation, we explore potential benefits and limitations of the models. Our evaluation is guided by good practices on empirical studies~\cite{Shull:2007} and is constraint to the following question:\\
\textbf{To what extent can models help comprehending the IEC 62443-4-1 security standard?} 


\subsection{Subject Selection}

As study environment we chose typical project settings at Siemens. There, product development projects follow the large-scale agile framework Scaled Agile Framework (SAFe) and the processes are deemed to comply with the 4-1 security standard. Most (Siemens) industrial projects that fit these characteristics demand direct collaboration between security experts and development teams. Moreover, development practitioners need to be more knowledgeable on security standards to accurately implement security requirements on a day-to-day basis. Therefore, we selected development practitioners working in these projects and security experts that join the projects on-demand.

Our selection comprises 16 experienced professionals. We defined profiles to distinguish their level of expertise according to their key role. \Cref{tab:6_evaluatorsPerProfileGroup} shows each role's background and the distribution in the interviews.

\begin{table}[htb]
    \centering
	\caption[Mapping of interviews to profile groups]{Mapping of interviews to profile groups with background. Adapted from \cite{Moyon:2018:thesis}}
	\scalebox{0.8}{
		\begin{tabular}{L{2cm}|C{1.5cm}|C{2cm}|L{10cm}}
			\cmidrule{1-4}    \multicolumn{1}{c|}{\textbf{Profile}} & \textbf{Sample size} & \textbf{Interview numbers} & \multicolumn{1}{c}{\textbf{Background}} \\
			\midrule
			Contributor IEC & 1 & 13 & \multicolumn{1}{l}{Industrial systems life cycle standardisation} \\
			\midrule
			Contributor SAFe & 1 & 12 & \multicolumn{1}{l}{Industrial systems agile development} \\
			\midrule
			Principal Expert & 3 & 4, 5, 8 & Industrial systems security standards and processes, secure design for industrial solutions \\
			\midrule
			Senior Expert & 4 & 1, 2, 6, 9 & Industrial systems cloud security, security processes improvement, IT security assessments \\
			\midrule
			Expert & 7 & 3, 7, 10, 11, 14, 15, 16 & Industrial systems agile development, quality compliance, design and development of access control systems, data privacy on smart cities, security design management, DevOps, security tools development, automated security testing, IT security in critical infrastructure \\
			\bottomrule
	\end{tabular}}%
	\label{tab:6_evaluatorsPerProfileGroup}%
\end{table}%

In this table, we distinguish between standards contributors and topic experts. We divided the 4-1 standard (\emph{Contributor IEC})and the contributos to the SAFe framework (\emph{Contributor SAFe}). The topic experts are categorized into \emph{Principle Experts} having broad knowledge and leading teams, \emph{Senior Experts} having deep knowledge and responsible for putting specific topics into practice, and general \emph{Experts}.

\begin{figure}[htb]
	\centering
	\includegraphics[width=0.7\textwidth]{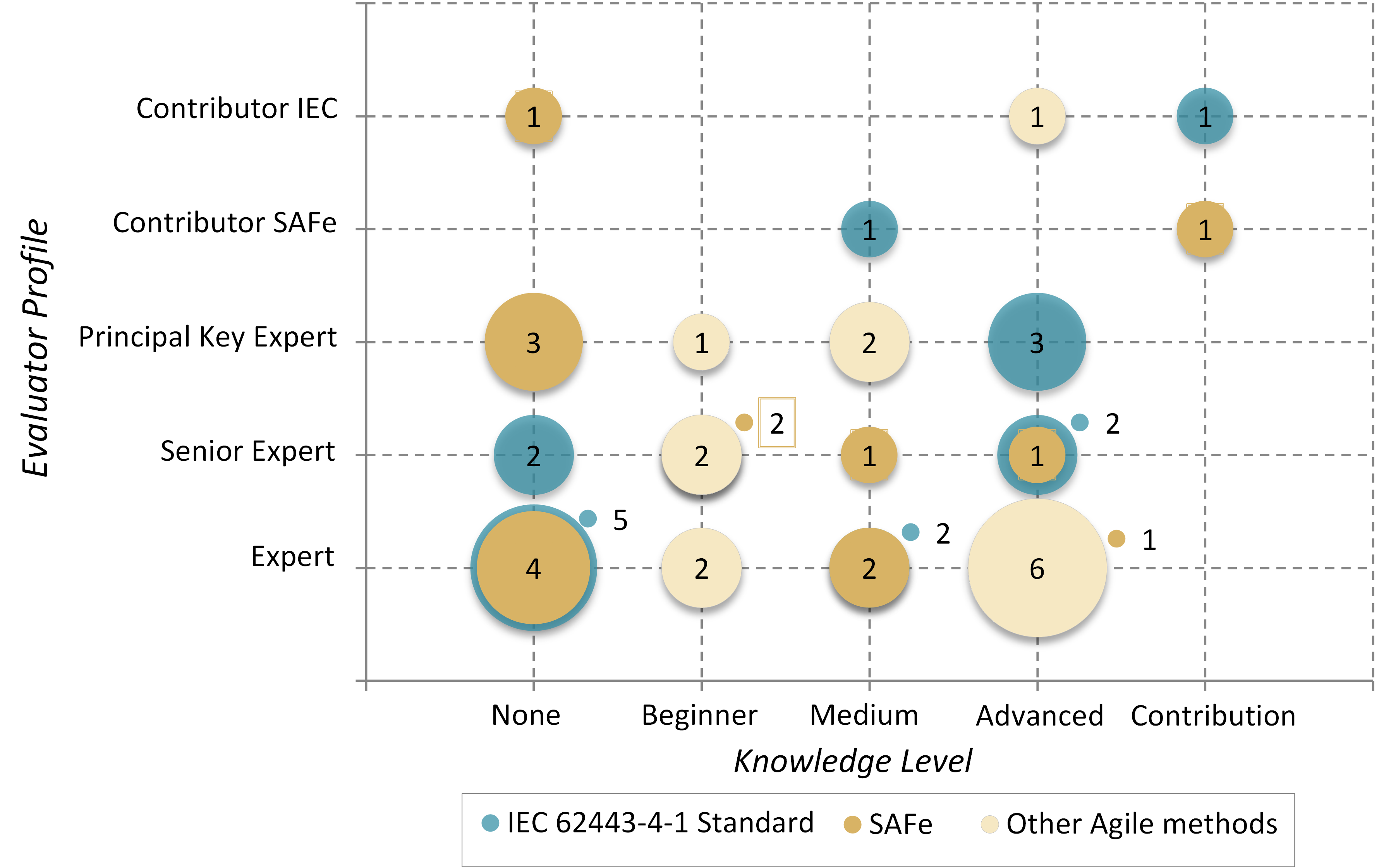}
	\small \caption[Subject knowledge of SAFe and 4-1]{Subject knowledge of IEC~62443-4-1 and SAFe or comparable process frameworks. Image adapted from \cite{Moyon:2018:thesis}}
	\label{fig:6_SAFeAnd41}
\end{figure}

Our subjects have different knowledge levels of the 4-1 standard and SAFe. \cref{fig:6_SAFeAnd41} shows the knowledge of our subjects in different areas, depending on their category. Even though not all of them know the 4-1 standard, all except for one are aware of other security standards such as ISO/IEC~27001 or other standards of the IEC~62443 family. Similarly, not all know SAFe, but all subjects are familiar with other agile process frameworks such as Scrum.

\vspace{-1pt}

\subsection{Survey Instrument}
Since our goal is to explore practitioners' opinions, we identified semi-structured interviews as the most suitable technique~(see \cite{Shull:2007}). Each interview lasted 1.5 to 2 hours and took place in an isolated environment with one interviewee and two interviewers. One interviewer actively followed a questionnaire and the other one documented the answers along the protocol. The questionnaire is available in the online material. 

\begin{figure}[htb]
	\centering
	\includegraphics[width=0.9\textwidth]{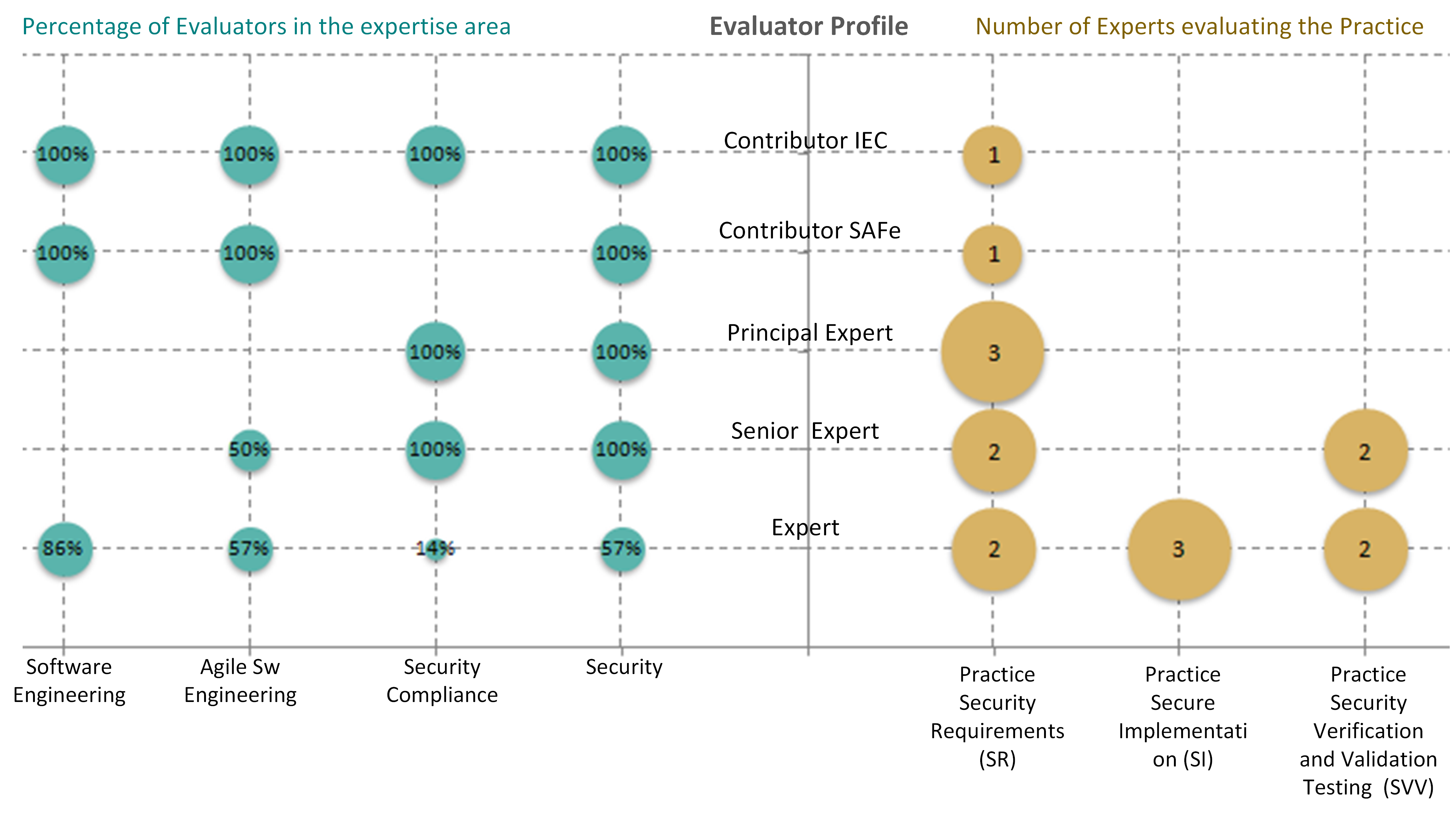}
	\small \caption{Distribution of IEC~62443-4-1 practices into profile groups. Right side: number of interviewees per practice. Left side: percentage of interviewees per expertise area. Image adapted from \cite{Moyon:2018:thesis}}
	\label{fig:6_expertiseAndSolution}
\end{figure}

Each interview deals with one 4-1 standard practice, which was chosen in advance according to the subject's background. Practices are either security requirements (SR), secure implementation (SI) or security verification and validation testing (SVV) as shown in \cref{fig:6_expertiseAndSolution}. 

Interview subjects intentionally received no instruction or training on the 4-1 models at all. They were first briefed about the interview flow and the purpose of the process models. Afterwards, we conducted the interviews in the following steps:

\begin{itemize}
    \item Interviewees were introduced to our 4-1 process models as well as their hierarchy (the entire overview models (see \cref{fig:41_global01}) and an individual practice model).
    \item They analyzed one of the 4-1 practice models without time limits (e.g. practice SR shown in \cref{fig:41_practice2}).
    \item They read a textual excerpt from the 4-1 standard corresponding to the practice.
    \item They answered the protocol questions about 4-1 (e.g. \quotes{Are process models easier to understand than the 4-1 text?}).
\end{itemize}

Finally, the interviewer notes on the questionnaire were discussed with the subject for clarification.

\subsection{Data Analysis}
Our preliminary evaluation is based on summarizing the answers to closed questions and clustering comments and concerns according to commonalities. We further analyzed the emphasis of answers to differentiate acceptance vs. conviction, rejection vs. repulsion, and neutrality vs. doubt. Hence, we tabulated answers according a 9-point Likert scale.

\section{Evaluation Results}
\label{sec:results}
The evaluation shows that understanding of the 4-1 standard is increased. 14 participants have claimed to understand our models better than the 4-1 standard. All subjects felt like Activities and Flows could be more easily identified, while 14 also identified artifacts more easily. Furthermore, all interviewees except for one claimed that they require less time for reading and understanding the model, when compared to the text. Moreover, 15 participants consider the models as useful for their work. 
Except for the required time to read, all subjects considered the model approach in all aspects as fitting at least equally well.

\begin{table}[htb]
	\centering
	\caption{Models common issues as referred by Interview Participants}
	\scalebox{0.8}{	\begin{tabular}{L{12cm}|L{2cm}|C{1.2cm}}
			\toprule
			\multicolumn{1}{c|}{\textbf{Evaluators' opinion}} & \multicolumn{1}{c|}{\textbf{Interviews}} & \multicolumn{1}{c}{\textbf{Total}} \\
			\midrule
			Find models helpful to have good overview of the standard. models don't  replace the document. Interviewees would use text and models. & 4, 5, 9, 11, 13, 14, 15, 16 & 8 \\
			\midrule
			Analyze the models and discuss doubts on the standard that they had beforehand & 5, 8, 11, 13, 14, 15 & 6 \\
			\midrule
			Would like to understand deep the notation, analyze in deep the elements and/or provide feedback to the notation & 2, 4, 8, 10, 11, 13 & 6 \\
			\midrule
			Analyze the models and ask where the standard says so. They consult the text and notice something new or another interpretation & 5, 6, 8, 11, 13 & 5 \\
			\midrule
			Provide feedback on visualization (line thickness, elements location)& 7, 8, 10, 13, 14 & 5 \\
			\midrule
			Have a  concern on models interpretation. Suggests PM approval and refinement by competent professionals e.g. IEC Committee  & 4, 6, 9, 11 & 4 \\
			\midrule
			Provide feedback on nomenclature & 4, 7, 8, 10 & 4 \\
			\midrule
			Have a concern about the suitable audience of the models & 5, 13, 15 & 3 \\
			\midrule
			Envision issues or feedback to the standard & 8, 13, 14 & 3 \\
			\midrule
			Requires to know content of artifacts, e.g. templates of documents & 1, 4, 5 & 3 \\
			\midrule
			Requires a catalog of models & 1, 12 & 2 \\
			\midrule
			Require clarification about the term artifact & 9, 11 & 2 \\
			\bottomrule
	\end{tabular}}
	\label{tab:6_clustersQ1}%
\end{table}%

In order to identify potential improvements, the comments provided by the subjects were clustered into categories. 
\Cref{tab:6_clustersQ1} summarizes common opinions on the model approach. Further outcomes of our study can be found below. For any specific participant statement, the respective interview of \autoref{tab:6_evaluatorsPerProfileGroup} is referenced:
\newline
\textbf{Security standard models are intuitive and precise.} Subjects read models quicker than the standard text and could also understand the global process better. Artifacts, activities, and flows are easier to identify than in text. They corroborated the criteria that the complexity could be reduced than when reading a standard: \quotes{\textit{I~can concentrate on specific things, that is why I like it}} (I14 - \textit{expert} security researcher). 
\newline
\textbf{Security standard models increase comprehension of the standard without replacing the text.} During the interviews, some respondents solved preexisting doubts while others rediscovered some statements of the standard, e.g., \textit{\quotes{Why there is no flow here?}} (I14 \textit{expert} product owner). Nonetheless, they recognized that a deep understanding of the standard requires to read the document, e.g., \textit{\quotes{text provides some reasoning that cannot fit in the diagram"} } (I4 security \textit{principal expert}). 
Also, the \textit{contributor IEC} (I13) appreciated the 4-1 standard  models. They started skeptic but ended up rather convinced. First, it took them time to read the 4-1 global model and analyze the detailed practice in depth. They asked for the reference of some elements in the model \textit{\quotes{Where does the standard say so?}}. Later, they discussed some model references to other members of the 62443 standard family. They agreed on some assumptions we had to make and openly discussed the accuracy of the standard referring to security requirements and its relationship to the threat model. They argued on the purpose of the corresponding text, thus, corroborating again the need of rich descriptions as a rationale. Finally, they approved our model while stating \textit{\quotes{[The] Model helps to see relationships, text has more description. We need both.}} 
\newline
\textbf{Security standard models support quick understanding.} Interviewees referred to different scenarios to apply models, such as (1) to report issues to the IEC committee, (12) to plan and estimate projects well, or (3) to solve issues related to knowledge transfer: \textit{\quotes{Models will help standardization steps in a memorable way.}} (I9 - security assessments \textit{senior expert}); \textit{\quotes{There were discussions within IEC about this reference to 4-2}} (I13 \textit{contributor IEC}); \textit{\quotes{I would like to use it in my lectures}} (I8 - security \textit{principal expert}). 
\newline
\textbf{Proficient review of security standard models motivates adoption.} Our models are a translation of the standard (as understood by us and validated by experts) to a visual language. Interviewees pointed out the risk of crossing the line between representation and interpretation/assumption. They would have felt more comfortable to follow the models after an accurate review and possible approval by the IEC Committee. However, the fact that the \textit{4-1 global PM} and \textit{practice 2} were approved by the IEC Contributor strengthens our confidence in the reliability and accuracy of our models. 
\newline
\textbf{Development Teams require derived models.} The \textit{IEC contributor} (I13), and a security \textit{principal expert} had concerns about confronting development teams with the 4-1 models. They thought teams would either underestimate the models or they might feel overwhelmed: \textit{\quotes{It is intimidating at first sight}}(I5). In addition, an expert system architect (I15) was highly skeptic when looking at the models: \quotes{\textit{My experience shows that models do not model reality, [...] I have not seen a process that is being followed}}. Even though at end of the interview, the subject (I15) describe to get more familiar with the standard, it is clear, more refinements of the models are needed for development teams. For this use case process modeling tools can be beneficial. 
\newline
\textbf{Artifact instances are needed.} Interviewees asked for examples of the artifacts. They consider models as a \quotes{to-do checklist}, therefore a template of the artifacts would help to estimate implementation effort. Also, it seems to be unclear what an artifact is according to the standard, e.g., \quotes{\textit{[...] standards don't say specifically these are artifacts [...] security requirements is not an individual document}} (I5 -\textit{principal expert}). Artifact instantiation is not trivial, for example for threat models, a main security analysis artifact, there is not yet a common agreement~(c.f.\cite{Shostack:2014,Fernandez:2016}). A future work goal is to explore artifacts instantiation.
\newline
\textbf{Security standard models catalog is needed.} Subjects reviewed parts of the models. Some of them had interest in a comprehensive list of all models to have an idea of the size. \quotes{\textit{To manage the catalog of flowcharts may be complicated}} (I5).
\newline
\textbf{Security standard models need precise notation for artifacts.} BPMN 2.0 provides limited elements to represent artifacts: document, data store, data message. These elements are not suitable to visually express the type artifact. We model most of our artifacts as documents, however, according to \textit{contributor IEC} \quotes{\textit{artifacts are not documents}} (I13). An evaluator would like to \quotes{\textit{Update notation to reflect type of artifact}} (I4).
\newline
\textbf{Security standard models require highlighting.} Evaluators gave feedback on how to improve models appearance, e.g., lines position, bold titles, and subtitles; e.g., \quotes{\textit{Change line from right to left}} (I8). 
\newline
\textbf{BPMN notation is intuitive.} Except for one, subjects read the models without introduction on BPMN. All of them understood models in limited time. Based on this observation, we perceive that \emph{BPMN is highly intuitive}. However, some elements of the last version BPMN 2.0 required clarification, e.g., event types escalation, signal, event trigger throwing and receiving. Also others were simply disregard, e.g., the collection symbol to describe a document artifact is a set, indirect flow. Some participants requested a legend for BPMN.


\section{Conclusion}
\label{sec:conclusion}
In this paper, we reported on our ongoing work towards using process models to implement and disseminate complex and easy to misunderstand security requirements derived from IEC~62443-4-1 standard. In scope of the paper at hand and of particular interest to us was to facilitate comprehensibility and focused discussions on the 4-1 standard for which we used BPMN process models. We evaluated this visualization by interviewing 16 industry experts.

Our results strengthen our confidence in that the models can be understood in a time-effective manner and challenge popular belief that especially agile processes are a \quotes{gateway to chaos} and, thus, not reconcilable with security and compliance concerns.
The unanimous response to our work was the exact opposite: Introducing large-scale agile processes demands a culture and mindset change. Even though not our primary intention, the models seemed to have helped to convey to skeptical practitioners that both secure and agile development is feasible at scale with reasonable effort. 

Our research indicates that models are an excellent way to mediate between agile practitioners and security experts. Particularly visual models allowed them to engage the challenge of continuous security compliance together. Moreover, these models pave the way for analyzing various further challenges of the research field: Do models increase the speed of adapting large organizations to secure agile processes at scale? Are models a better way of getting security norms accepted into daily software engineering activities? Can models provide guided and precise support for secure agile security governance? We are confident that our contribution supports researchers to further investigate these questions.



\vspace{-1pt}
\subsection{Relation to Existing Evidence}

Our study is in tune with existing trends of empirical studies on secure software engineering~\cite{Othmane:2017}, but extends the study population in number and profile. To the best of our knowledge, preceding studies involved up to 11 practitioners with mixed background or students as subjects and focused on valuated, yet isolated topics. An integrated view on security standard compliant scalable agile framework was not in their scope. Our contribution is aimed at this gap and involves 16 experienced professionals, partially with contributing roles to the standards or decision-making roles in the organization. We focused on the highest ranking experts available. Their opinion is the closest to certainty in a timely evaluation. 
\vspace{-1pt}
\subsection{Limitations and Threats to Validity}
\label{sec:evaLimitations}

Qualitative studies inherently carry limitations and interview research in particular has threats to validity that need discussion, the most important of which shall be discussed here.

The individual expertise of each participant might influence their attention and interpretation of security requirements as well as agile practices captured in the models. We tried to mitigate this with discussion-intensive preparation procedures but also by letting subjects interpret the models as they are without any further instruction. We were interested in potential bias towards the subject of security compliance as that reflects on the projects where those models shall be applied.

Similarly, involving experts from each respective field carries the risk of self-selection and confirmation bias. To mitigate this we selected subjects according to typical roles in the target organization environment instead of their particular interest in the topic. The same is true for which part of the 4-1 standard they reviewed (requirements, implementation, or testing). We also designed interview plan and questionnaire accordingly and allocated interviewees to models based on previously defined profiles.

\vspace{20pt}
\begingroup
\let\clearpage\relax
\bibliographystyle{splncs04}
\bibliography{bibliography.bib}
 \endgroup
\end{document}